\documentclass{aa}
\usepackage{graphicx}

\begin{document}

\title{Identification of 13 Cepheids and 333 Other Variables in M31}

\author{Y. C. Joshi\inst{1}, A. K. Pandey\inst{1}, D. Narasimha\inst{2},
R. Sagar\inst{1} and Y. Giraud-H\'{e}raud\inst{3}}

\offprints{Y. C. Joshi, \\
           \email{yogesh@upso.ernet.in}
            }

\institute{ State Observatory, Manora peak, Naini Tal - 263129, Uttaranchal, 
India 
          \and
              Tata Institute of Fundamental Research, Homi Bhabha Road, 
Mumbai -- 400 005, India
           \and
              Laboratoire de Physique Corpusculaire, College de France, 
Laboratoire associe au CNRS-IN2P3 (URA 6411), 11 place Marcelin Berthelot, 
75231 Paris Cedex 05, France
              }

\date{Received ------ /accepted ---------}

\abstract{
We present Cousins $R$ and $I$ band photometry of variable stars in a
$\sim13'\times 13'$ region in the disk of M31 galaxy, obtained during 141
nights. Of the 26 Cepheid variables present in the region, two are newly
discovered, 11 are classified as Cepheids for the first time and 13 are
confirmed as Cepheids. The extensive photometry of these Cepheids enabled
us to determine precise phase and amplitude of pulsation which ranges from
0.11 to 0.48 mag in $R$ band. The period of variability ranges from $\sim$7.5
to  56 days. The period-luminosity diagram is used to derive a distance
modulus of 24.49$\pm$0.11 mag for M31 galaxy. We also report variability in
333 other stars, of them, 115 stars appear to be long period variables,
2 suspected eclipsing binaries and remaining 216 are irregular variables.
}
\authorrunning{Y. C. Joshi et al.}
\titlerunning{Identification of 13 Cepheids and 301 New Variables in M31}
\maketitle
\section{Introduction}
In recent years, the Andromeda galaxy (M31) has been a target of search
for gravitational microlensing events (Crotts \& Tomaney 1996, Ansari et al.
1997). To detect lensing events, continuous observations are needed for 
a long duration though for a short period of time each night. Such 
observations are, therefore, very useful to search for variable stars 
(e.g. Cepheids, Miras) also. 
The catalogues of variable stars compiled from such a monitoring programme
are generally complete within the flux limitation,
because of the continuous observation of the same field over a long period.
Various groups (e.g. MACHO, EROS, OGLE) dedicated to 
search for microlensing events in the Galactic Bulge and Magellanic Clouds 
have already identified a large number of variable stars as a 
bi-product (Welch et al. 1997, Udalski et al. 1999a) and 
the catalogues of Cepheid variables brought out by these projects
(Beaulieu et al., 1995, Udalski et al. 1999b)
have given insight into the pulsation properties of Cepheids.

In collaboration with the AGAPE (Andromeda Gravitational Amplification Pixel 
Experiment) group, we started Cousins $R$ and $I$ photometric observations
of M31 in 1998 with an aim to search for microlensing events.
Based on the observations of first 2 years, Joshi et al. (2001, hereafter
referred as paper I) reported 8 variables, 7 of them were suspected as
Cepheids and one Mira-like variable. In this paper, we extended our
analysis to detect the variable stars using four years of 
observations. Details of our observations are given in next section while 
Sect. 3 outlines the detailed photometry of our data. The detection of
variable stars is described in Sect. 4 which leads to discovery of
some new
Cepheids. The catalogue of Cepheids, their period-luminosity relation,
colour-magnitude diagram along with a brief discussion of some of the
Cepheids are presented in Sect. 5. Other type of variables along with
discussion and conclusion are given in the remaining part of the paper.
\begin{table*}[ht]
{\bf Table 1}
{Characteristic parameters of the two CCDs used for observations.}

\centering
\begin{tabular}{clccccc}       
\hline
Size of CCD        &  Field             & \multicolumn{2}{c} {Quantum Efficiency} 
& Gain & Readout Noise & Pixel size \\
(pixel$^2$)        &  (arcmin$^2$)      &    R(\%)   & I(\%)  & (e$^{-}$/ADU) 
& ($e^{-}$) & (arcsec)    \\ \hline
$1024 \times 1024$ & $\sim6 \times 6$  &   35 & 31  & 2.96 & 4.1 & $\sim$0.37 \\
$2048 \times 2048$ & $\sim13\times 13$ &   74 & 61  &10.00 & 5.3 & $\sim$0.37 \\\hline
\end{tabular}
\end{table*}
\section{Observation and image processing}
The Cousins $R$ and $I$ broad band photometric observations were carried out 
at the f/13 Cassegrain 104-cm Sampurnanand Telescope of the State Observatory, 
Naini Tal. 
The CCD observations of the M31 were started in November 1998
using a $1024 \times 1024$ pixel$^{2}$ CCD covering $\sim6'\times 6'$ field.
A large size CCD containing $2048 \times 2048$ pixel$^{2}$ covering an area
of $\sim13'\times 13'$ was used for the observations in later years. The
target field ($\alpha _{2000}$ = $0^{h} 43^{m} 38^{s}$
and $\delta_{2000}$ = $+41^{\circ}09^{\prime}.1$) is centered at a distance
of about 15 arcmin away from the center of M31.
The detailed overview of telescope and instruments are given in 
paper I. A brief summary of the characteristic parameters of the 
two CCDs used for observations are given in Table 1.
To minimize air mass effects, most of the observations were taken when zenith
distance was $\le$ 3 hour. We have accumulated 468 and 383 data points
in $R$ and $I$ filters respectively during 141 observational nights spanning
over $\sim$1200 days. Table 2, available electronically either from authors or
at the CDS, lists the date of observation, corresponding Julian date and total
exposure time. The average seeing during the entire observing runs was $\sim$2
arcsec.

Data reduction have been performed using the MIDAS and IRAF softwares.
Preliminary steps of the image processing include bias and flat fielding
corrections. As the dark current during the maximum exposure time of a frame
is negligible, it has not been corrected. The cosmic rays contaminations
were removed independently from each frame.

The observations were performed in unbinned (0.37$\times$0.37 arcsec$^2$) as
well as in 2$\times$2 binned (0.74$\times$0.74 arcsec$^2$) mode during 
different runs of observations. As the main aim of our
observations is to search microlensing events in a combined data of Naini Tal
(NTL) and AGAPE (observed with 1.3m MDM telescope at Kitt peak, Arizona, a
detailed discussions is given in Calchi Novati et al. 2002), we
aligned all the NTL images with respect to a MDM reference image.
The NTL images were scaled down or up to MDM pixel size (0.5$\times$0.5
arcsec$^2$). These aligned images, in combination with AGAPE observations,
will be used to search for microlensing events using the pixel method
described in detail by Ansari et al. (1997). 
\begin{figure}[h]
\centering
\includegraphics[height=12.0cm,width=9.0cm]{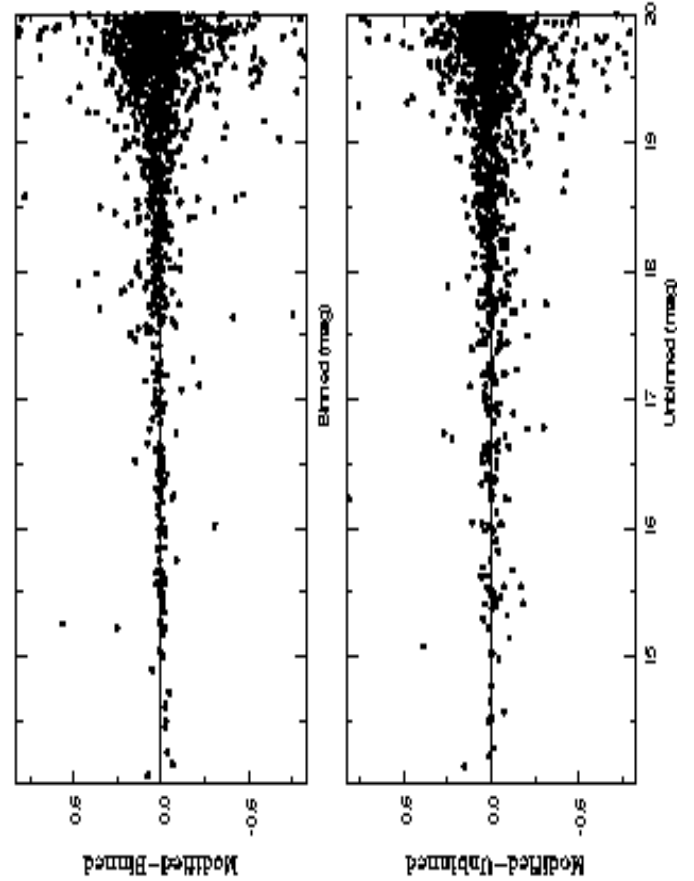}
\caption{The plot shows difference in photometries carried out on original
and modified images as a function of magnitude. In the upper panel original
pixel size (0.74 arcsec) was scaled down (0.5 arcsec), whereas in the lower
panel the pixel size (0.37 arcsec) was scaled up (0.5 arcsec).}
\end{figure}
\section{Photometry}
To check whether profile fitting photometry could be carried out on the
modified NTL images, we compared the photometries carried out on original
image and on modified images and the results of a particular night are
displayed in Fig. 1. The comparison shows good agreement between the magnitude
derived from original and modified images. Therefore we carried
out photometry on the modified images. In order to increase signal to noise
ratio, all the images of a night were stacked together resulting one frame per
filter per night. This then provides us 133 $R$ and 166 $I$ data points which
were used in further analysis. 

Using DAOPHOT `FIND' routine, we identified $\sim$4400 resolved stars in the 
reference frame at $4\sigma$ detection level. Stellar photometry for all the 
images in both filters has been carried out for these resolved stars in 
`fixed-position mode' using DAOPHOT photometry as described by 
Stetson (1987). PSF was obtained for each frame using 
25-30 uncontaminated stars. DAOPHOT/ALLSTAR (Stetson 1987) routine was used 
to calculate the instrumental magnitude of the detected stars for each 
individual frame. The internal errors estimated from the 
S/N ratio of the stars as output of the ALLSTAR are given in Table 3 as a 
function of brightness. The error become large ($>0.1$ mag) for stars 
fainter than $R \sim$20.0 mag. 
\begin{table}[h]
{\bf Table 3}
{Internal photometric errors as a function of brightness. $\sigma$ is
the standard deviation per observation in magnitude.}

\centering
\begin{tabular}{cccccc}
\hline
Magnitude range& $\sigma$$_{R}$ & $\sigma$$_{I}$\\
\hline
$\le$14.0  &0.01&0.01\\
14.0 - 15.0&0.01&0.01\\
15.0 - 16.0&0.01&0.01\\
16.0 - 17.0&0.01&0.02\\
17.0 - 18.0&0.02&0.04\\
18.0 - 19.0&0.04&0.06\\
19.0 - 20.0&0.07&0.13\\
20.0 - 21.0&0.15&0.33\\
\hline
\end{tabular}
\end{table}
\subsection {Photometric Calibration}
The absolute calibration has been done using Landolt's (1992) standard field 
SA98 observed on a good photometric night of 25/26 October, 2000. The airmass
ranges from 1.3 to 2.1 during the observations. Atmospheric extinction
coefficients determined from these observations are 0.11$\pm$0.01 and
0.08$\pm$0.01 mag/airmass in $R$ and $I$ filters and they have been used in
further analysis. Thirteen standard stars having range in brightness 
$(11.95 < V < 15.84)$ and colour $(0.09 < (R-I) < 1.00)$ were used to
derive following transformation equations.
\begin{equation}
\Delta(R-I) = (0.96 \pm 0.01) \times \Delta(r-i)
\end{equation}
\begin{equation}
\Delta(R-r) = (0.03 \pm 0.03) \times (R-I)
\end{equation}
where $R$ and $I$ are the Landolt standard magnitudes while $r$ and $i$ 
are the corresponding instrumental magnitudes. The zero point errors are
about 0.02 mag in $R$ and 0.01 mag in $(R-I)$. Equations (1) and (2) were
used to generate 50 secondary standards in the target field observed on
the same night by accounting the differences in exposure times and
air-masses. To standardize the remaining stars, differential photometry has
been performed with these secondary stars rejecting those which were showing
more than $3\sigma$ deviation. This process yields an accuracy of 0.03 mag in
zero points. A variation of $\pm0.1$ mag was found around the mean value in
the secondary stars itself during the whole observing runs which can
be treated as an accuracy in NTL photometry. 
\subsection {Comparison with the previous photometries}
A comparison of the present $I$ band data with those available in the
literature (Magnier et al. 1992, Mochejska et al. 2001 for the DIRECT
collaboration) has been shown in Fig. 2 as it is the only common filter. An
offset of 0.13 mag is observed between our and Mochejska et al. (2001) $I$
band photometry while a difference with Magnier et al. (1992) data indicates
a magnitude dependence with a slope of 0.02. A similar slope is seen between
Mochejska (2001) and Magnier et al. (1992) data. These results suggest that
Magnier et al. (1992) data is colour dependent while Mochejska et al. (2001)
data has zero point offset.
\begin{figure}
\centering
\includegraphics[height=12.0cm,width=9.0cm]{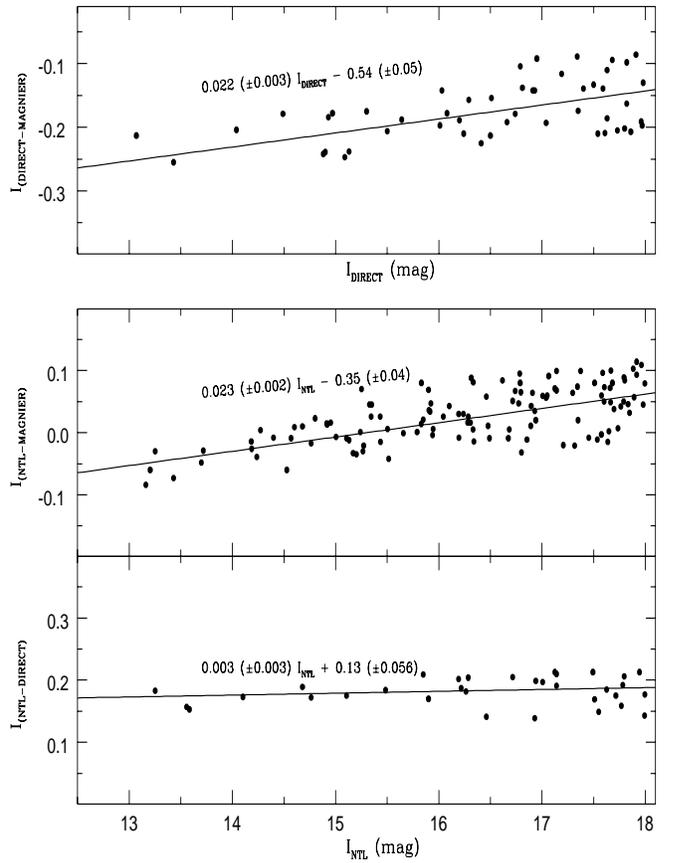}
\caption{Comparison of the present $I$ band CCD photometry with those given by
Magnier et al. (1992) and Mochejska et al. (2001).}
\end{figure}
\section {Detection of variable stars}
\subsection{Selection Criteria}
We do not have all the observations in good photometric sky conditions.
Therefore we have rejected all those data points showing photometric error of 
$>$ 0.2 mag. Further, for each star, the average DAOPHOT error and its standard 
deviation was calculated using its observations on different nights.
Measurements with errors 
exceeding to average error by more than 3$\sigma$ were flagged as `bad' and 
removed from further analysis. The whole procedure was iterated thrice. 
Normally 1 to 10 points were removed in this process. Because two different 
CCDs with different quantum efficiencies (see Table 1) and different exposure
times (see Table 2) were used, limiting magnitudes were different in different
years. Consequently some stars could not be identified in one or other years of 
observations. All the stars were not present in every frame due to different
observing conditions and CCD orientations on various nights. In order to derive 
useful results, we consider only those stars in our further analysis which have 
more than 40 $R$ data points. Stars which were showing more than $5\sigma$
variation in 3 consecutive data points are searched for variability. 
Only $R$ band photometric data is used for this as the data sample in $R$ 
filter (133) is larger than that in $I$ filter (116). Also photometric errors
at a given brightness are generally less in $R$ filter (see Table 3). Finally
we analysed the variability in these stars explicitly by visual monitoring and
also using $I$ filter data. In this way, we detected 359 variable stars in our
field. 
\subsection {Period determination}
To find periods from unequally spaced data, we used a modified version of the
period-searching program by Press and Rybici (1989) based on the method of
Horne \& Baliunas (1986). The data were initially phased 
for all periods between 5 and 600 days searched in an increment step of 0.6
day. To further refine the period, an increment of 0.1 day was used around
thus derived period. In this way, we could determine period of 141 variables.
The remaining stars are either non-periodic or long-period variables.
\subsection {Mean magnitude} We calculated phase weighted apparent mean
magnitude for all the Cepheid variables as suggested by Saha et al. (1994)
$$ \overline m = -2.5\log_{10} \mathrm{\sum_{i=1} ^{n}}\ 0.5
(\phi_{i+1}-\phi_{i-1}) 10^{-0.4 m_{i}} $$
where n is the total number of observations, $\phi_{i}$ is the phase of 
$i^{th}$ observation in order of increasing phase after folding the period. 
The equation requires non-existent entities $\phi_{0}$ and $\phi_{n+1}$ 
which is set identical to $\phi_{n}$ and $\phi_{1}$ respectively. The 
estimation of mean magnitude by the phase-weighted method is superior to 
an ordinary mean, which minimizes the systematic biases from loss of faint 
measurements in the mean magnitude (Saha \& Hoessel 1990). For other variables,
mean magnitude is estimated simply by intensity averaging of all the data
points.
\subsection {Astrometry} 
Transformation equations were derived to convert pixel 
coordinates (X,Y) into celestial coordinates ($\alpha_{2000}$, $\delta_{2000}$) 
using 324 reference star positions from the USNO\footnote {United State Naval 
Observatory} catalogue. These coordinates agree within $\sim$0.1 arcsec with
those given in Magnier catalogue (Magnier et al. 1992).
\section {Cepheid Variables}
Cepheid variables are post-main 
sequence stars occupying the instability strip in H-R diagram. Their light
curves display the characteristic `sawtooth' pattern, with periods ranging
from a few days to more than 100 days. We have identified 26 Cepheid variables
in the disk of M31 (Fig. 3). The Cepheids were identified by looking for the
stars with light curves similar in appearances to known Cepheids of M31. The
period of the Cepheids was determined as described in Sect. 4.2. The periods
estimated independently in two filters agree very well for most of the Cepheids;
however difference in the period exists for  a few short period variables due
to their low amplitude as well as comparatively larger photometric error in
$I$-band. Therefore, the period calculated using $R$ filter data was considered 
as the period of the star and phase in both $R$ and $I$ filters have been
evaluated using this period.

Since most of the Cepheids are believed to be in their second crossing of the
instability strip, their period of pulsation is directly related to the stellar 
mass, and hence, to the main sequence life time of the star. Hence a rough
estimate of the age of the star can be obtained from the observed pulsation
period of the Cepheids. We have used the following relation given by Magnier
et al. (1997b) to estimate the age of Cepheids:
\begin{equation}
$logA$ = 8.4 - 0.6 $logP$
\end{equation}
where $A$ is the age of the Cepheid in years and $P$ is the pulsation period in 
days. As Cepheids detected in our study have period ranging from $\sim$7.5 to 
56 days, the age of the Cepheids varies from $\sim$22 to 
75 Myrs with a peak of 48 Myrs. Though there is an uncertainty of 
$\pm$ 0.15 in log$A$ (Magnier et al. 1997b), this relation provides an 
indicative age of the Cepheids in M31, and hence some idea
of the recent star formation history in the field of our observations.
\begin{figure}
\centering
\includegraphics[height=9.0cm,width=9.0cm]{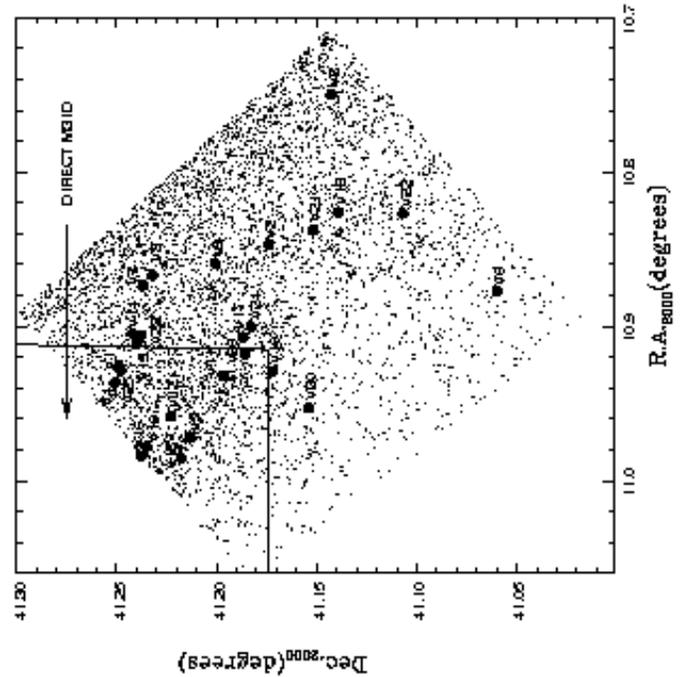}
\caption{Location of the 26 Cepheids identified in our data juxtaposed over 
the entire observed field of M31. Small dots indicates the position 
of 4400 stars observed in our field. Overlapping area of DIRECT M31D field
is also shown.}
\end{figure}

In Table 4, we present characteristics of these 26 Cepheid variables detected
in our study. The identification number of the star, celestial coordinate for
J2000, phase weighted mean magnitude in both $R$ and $I$ filters, amplitude of
variability in $R$ filter, period, age of the Cepheid (estimated using Eq. 3)
and total number of points in $R$ filter used for the light curve study are
given. The Cepheids are sorted in the increasing order of their periods. The
reference to variability reported in the literature and corresponding period,
if known, are listed in the last two columns. Out of the catalogued 26 Cepheids,
the variability is being reported for the first time for stars  V22 and V24;
moreover 13 of the variables are identified as Cepheids for the first time.
\begin{figure*}
\centering
\includegraphics[height=23.0cm,width=18cm]{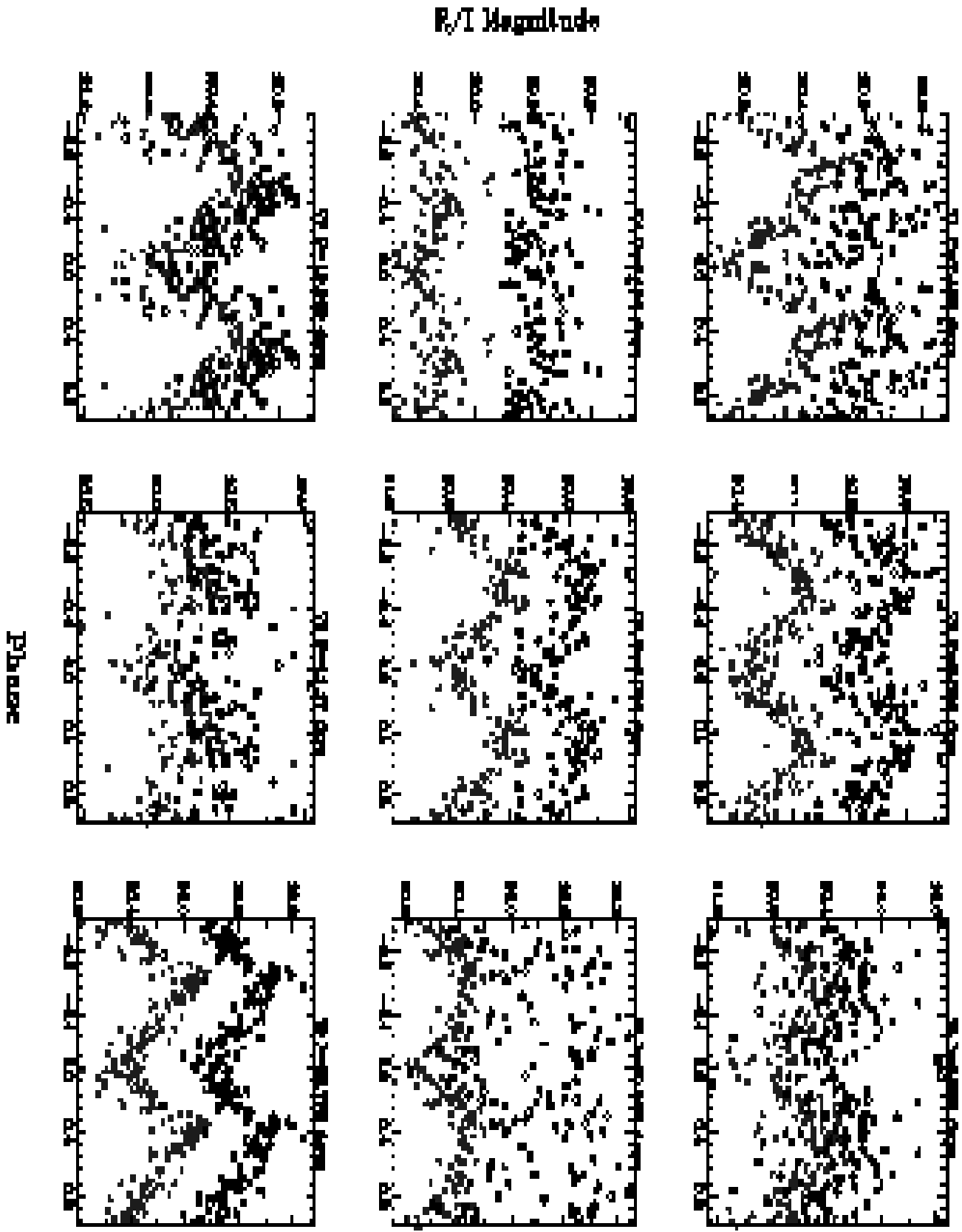}
\caption{Phase-magnitude diagram is plotted for 26 Cepheids detected in the
present study. Filled and open circles represents $R$ and $I$
magnitudes respectively.} 
\end{figure*}
\begin{figure*}
\centering
\includegraphics[height=23.0cm,width=18cm]{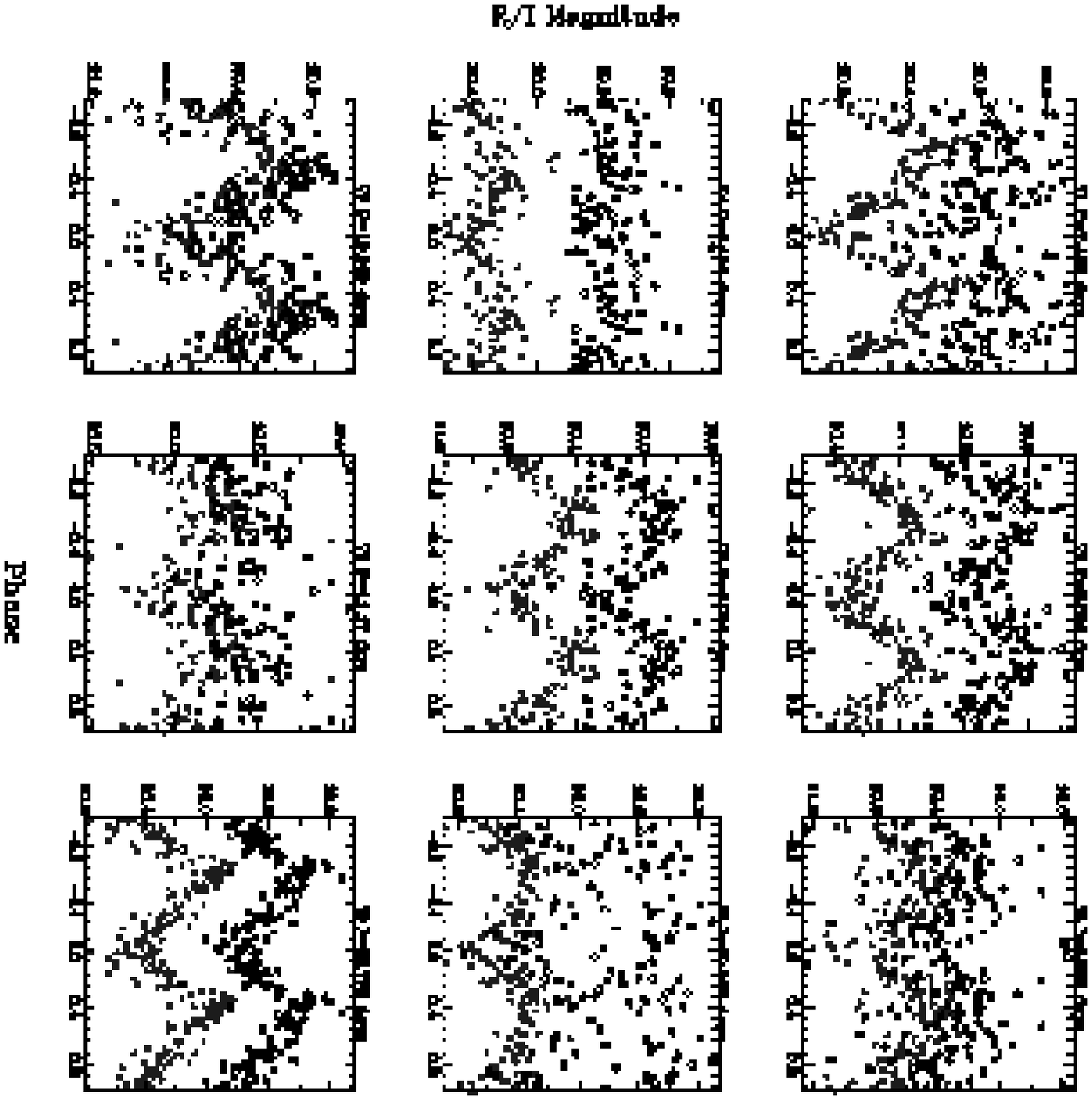}
continued...
\end{figure*}
\begin{figure*}{'}
\centering
\includegraphics[height=23.0cm,width=18cm]{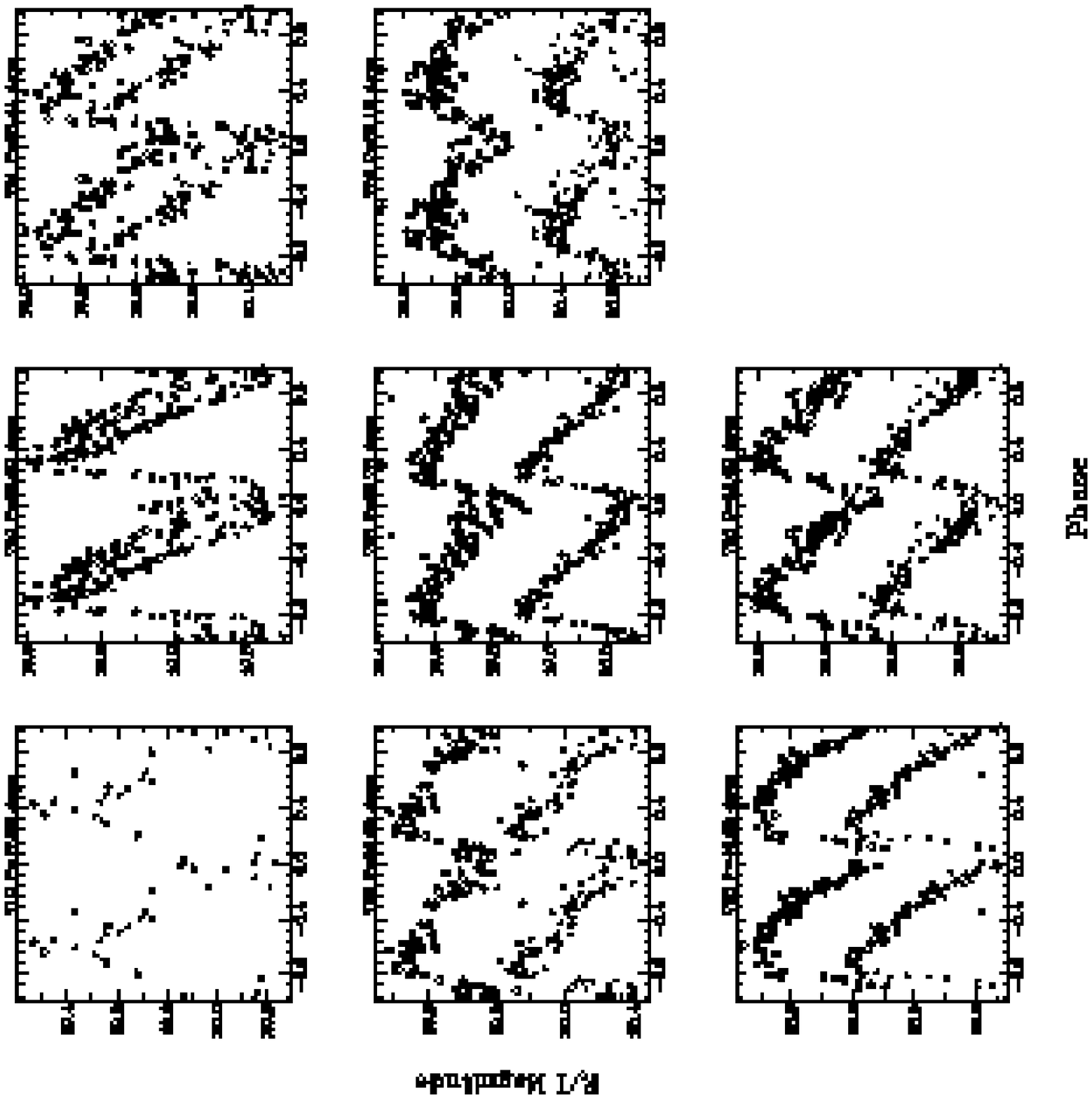}
continued...
\end{figure*}
\begin{table*}[ht]
{\bf Tbale 4.}
{A list of 26 Cepheids observed in our study with their characteristic
parameters. Star identification by KAL99, Tomaney \& Crotts (1996), MAG97 and
Berkhuijsen, E. M. et al. (1988) are prefixed with K, TC, M and B respectively
in column 10. The periods of the 13 Cepheids given in the catalogues are given
in the last column. The Cepheids discovered and classified by us are marked
as $\dag$ and $*$ respectively in the first column.}

\begin {center}
\begin{tabular}{ccccccccccc}
\hline
&    &   &    &  &  &    &  &  & & \\
Star & $\alpha$ &  $\delta$ & $\overline R$   & $\overline I$  & 
 $\Delta_R$    &    Period & Age & N  &  Other & Period\\ 
ID & (deg)    &  (deg)  &  (mag)  &  (mag)  & (mag) & (days)  & (Myrs)  & & Identification             & (days) \\ \hline
 V1 & 10.9321 & 41.1970 & 20.48 & 19.98 & 0.27 &  7.459$\pm$0.002 &  75 & 123 & K V883  & 7.459 \\          
 V2$^*$ & 10.8469 & 41.1737 & 20.17 & 19.69 & 0.15 &  8.566$\pm$0.003 &  69 & 124 & TC 170  &  ---  \\          
 V3$^*$ & 10.8669 & 41.2320 & 20.61 & 20.28 & 0.22 &  8.836$\pm$0.004 &  68 & 120 & TC 18   & ---   \\          
 V4 & 10.9366 & 41.2503 & 20.28 & 19.54 & 0.11 &  9.160$\pm$0.008 &  67 &  86 & K V1219 & 9.173 \\          
 V5 & 10.9721 & 41.2128 & 20.56 & 20.04 & 0.19 &  9.790$\pm$0.005 &  64 &  92 & K V2879 & 9.790 \\          
 V6$^*$ & 10.8770 & 41.0601 & 20.43 & 19.76 & 0.15 & 10.383$\pm$0.009 &  62 &  93 & TC 76   &  ---  \\          
 V7$^*$ & 10.8736 & 41.2367 & 20.42 & 20.27 & 0.28 & 10.500$\pm$0.004 &  61 & 123 & TC 16   &  ---  \\          
 V8 & 10.7500 & 41.1426 & 19.89 & 19.55 & 0.17 & 11.17 $\pm$0.01  &  59 &  95 & M 65    & 25.0$\pm$5.0\\ 
V9$^*$ & 10.8594 & 41.2004 & 20.21 & 19.60 & 0.26 & 13.773$\pm$0.006 &  52 & 126 & TC 20   &  ---  \\          
V10$^*$ & 10.9290 & 41.1715 & 20.77 & 19.84 & 0.48 & 14.420$\pm$0.006 &  51 & 116 & TC 85   &  ---  \\          
V11 & 10.9255 & 41.2489 & 19.57 & 18.87 & 0.16 & 15.26 $\pm$0.01  &  49 &  96 & K V635  & 15.255\\          
V12 & 10.9576 & 41.2227 & 20.84 & 20.08 & 0.32 & 15.46 $\pm$0.01  &  49 &  89 & K V2286 & 15.464\\
V13 & 10.8259 & 41.1386 & 19.82 & 19.46 & 0.40 & 15.76 $\pm$0.01  &  48 &  94 & M 68    & 14.0$\pm$2.8\\ 
V14$^*$ & 10.9049 & 41.2419 & 19.93 & 19.58 & 0.22 & 15.90 $\pm$0.01  &  48 & 121 & TC 194  &  ---  \\          
V15$^*$ & 10.9115 & 41.2396 & 20.79 & 19.91 & 0.30 & 15.95 $\pm$0.01  &  48 & 126 & TC 196  &  ---  \\          
V16 & 10.9775 & 41.2348 & 20.28 & 19.74 & 0.40 & 16.38 $\pm$0.02  &  47 &  47 & K V3198 & 16.345\\          
V17$^*$ & 10.9069 & 41.1868 & 20.12 & 19.60 & 0.39 & 16.60 $\pm$0.01  &  47 & 124 & B 4614  &  ---  \\          
V18 & 10.9853 & 41.2176 & 19.47 & 19.09 & 0.21 & 17.73 $\pm$0.01  &  45 &  91 & K V3583 & 17.703\\          
V19 & 10.9839 & 41.2374 & 19.83 & 19.60 & 0.32 & 17.83 $\pm$0.03  &  45 &  18 & K V3551 & 16.699\\          
V20$^*$ & 10.9526 & 41.1540 & 19.20 & 18.99 & 0.35 & 20.09 $\pm$0.01  &  42 &  96 & TC 207  &  ---  \\          
V21 & 10.8379 & 41.1514 & 19.74 & 19.31 & 0.39 & 21.44 $\pm$0.02  &  40 &  96 & M 69    & 13.0$\pm$2.6\\ 
V22$^\dag$ & 10.8272 & 41.1071 & 20.01 & 19.19 & 0.29 & 26.99 $\pm$0.04  &  35 &  92 & ---     &  ---  \\          
V23$^*$ & 10.9059 & 41.2379 & 19.78 & 18.92 & 0.34 & 28.78 $\pm$0.02  &  33 & 127 & TC 30   &  ---   \\         
V24$^\dag$ & 10.9002 & 41.1823 & 20.55 & 19.57 & 0.23 & 35.12 $\pm$0.05  &  30 & 121 & ---     &  ---   \\         
V25 & 10.9293 & 41.2475 & 18.89 & 18.35 & 0.31 & 43.53 $\pm$0.08  &  26 &  97 & K V836  & 43.371\\          
V26 & 10.9183 & 41.1856 & 19.36 & 18.82 & 0.22 & 56.02 $\pm$0.08  &  22 & 124 & K V164  & 56.116\\ \hline
\end{tabular}
\end{center}
\end{table*}
The remaining 13 Cepheids have been known from earlier studies. The present
data confirms their variability and our periods are generally in good
agreement with the reported value.

In Fig. 4, we show the phase-magnitude diagram for all the 26 Cepheids in
both $R$ and $I$ filters. The $I$-band light curves show more scattering
particularly for short period and low amplitude Cepheids. This could arise
due to the following reason: the amplitude of Cepheids decreases with increasing
wavelength (e.g. Freedman et al. 1985) and Freedman (1988) found a
ratio of 1.00:0.67:0.44:0.34 in amplitudes of $B$:$V$:$R$:$I$ filters. As
most of these Cepheids have pulsation of $\sim$0.1-0.2 magnitude in $R$
filter, the amplitude of pulsation in $I$ filter is even less and becomes
comparable to the photometric errors.

\subsection {Comparison with earlier studies}
Our target field in M31 has been observed only in a few earlier surveys. The 
$V$ band photometric observations have been obtained by Magnier et al.
(1997a, hereafter referred to as MAG97) for about 25 days. They have observed
9 fields of M31 and each field covered a region of $\sim$11 $\times$ 11
arcmin$^{2}$ on the sky. They found 130 Cepheids in the survey. As we have
observed smaller field, only 3 of them are located in our field and we have
identified all of them. In Table 4, we have compared period of these stars with 
those found in our data set. One can easily notice a large discrepancy between
the periods obtained in two surveys which we attribute to insufficient
observations by MAG97.

$BVI$ photometry has been carried out by Kaluzny et al. (1999, hereafter
referred to as KAL99) under the DIRECT project. Out of 35 Cepheids detected
in DIRECT project in the field M31D, 11 are located in our field of
observations. We have identified 10 of them and notice that except in the case
of V19, the periods determined in the two surveys are in excellent agreement.
The discrepancy in the period of V19 appears to be due to inadequate data in
the present study (see Table 4). One of the Cepheid (K V952) reported in KAL99
could not be detected in our data due to blending effect as this star is highly 
blended by a neighbouring bright red star. 

For Cepheids V7, V9, V14, V15, V17, V23 and V26, the parameters given in Table
4 supersedes our earlier results given in paper I, as these are derived using
data with much longer temporal coverage.
\subsection{The Period-Luminosity (P-L) Diagram}
The Cepheid variables exhibit a stable periodic variability and there is an
excellent correlation between their mean intrinsic brightness and pulsation
period. Consequently, the Cepheid Period -- Luminosity  relation provide an 
important standard candles to measure distances to galaxies up to the Virgo
Cluster, by comparison of their absolute magnitudes inferred from P-L relation
with their observed apparent magnitudes. Based on the study of Cepheids located 
in LMC, SMC and IC 1613 galaxies, Udalski et al. (2001) found that the zero
points of the P-L relations have no significant dependence on metallicity. In
fact, they are constant to within $\sim$0.03 mag. However, the values could be
affected due to other effects, e.g. blending, which can cause up to 10\%
uncertainty in the distance determination of M31 (Mochejska et al. 2000). We
have therefore used metallicity independent P-L relation in the present
analysis.

Using a set of 32 LMC Cepheid variables, Madore \& Freedman (1991) have 
obtained an equation of the ridge line in P-L diagram for Cousin $R$
filter as:
\begin{equation}
M_{R} = -2.94(\pm 0.09)(log$P$-1) - 4.52(\pm 0.04)
\end{equation}
while using a large sample of 658 Cepheids, Udalski et al. (1999c)
derived a PL relation for Cousin $I$ filter as:
\begin{equation}
M_{I} = -2.96(\pm 0.02)(log$P$-1) - 4.90(\pm 0.01)
\end{equation}
where we adopt a true distance modulus of 18.5 mag for the LMC (Freedman et al.
2001) instead of 18.2 mag used by Udalski et al. (1999c).
\begin{figure}[h]
\centering
\includegraphics[height=12.0cm,width=9.0cm]{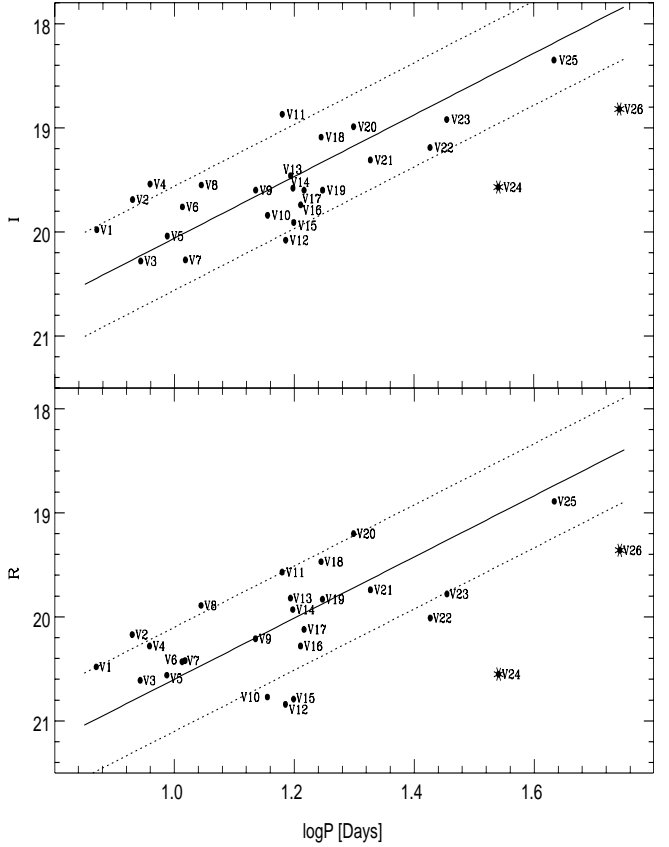}
\caption{Period-Luminosity relation for 26 Cepheids observed in our field.
Lower panel indicates P-L relation for $R$ filter while upper panel
indicates the same for $I$ filter. Filled circle: fundamental mode Cepheids; 
asterisk: possible Population II Cepheids. No corrections for reddening have
been applied to either data set. The slope of the fitted straight lines are
fixed at dm/d log$P$ = $-2.94$ and $-2.96$ for $R$ and $I$ filters respectively.
The dashed envelope lines have been drawn 0.5 magnitude
from the straight line fit (see text for details).} 
\end{figure}

The apparent mean magnitudes of 26 Cepheids are plotted as a function of logP
in Fig. 5. The slope of the straight line is fixed as $-2.94$ and $-2.96$ for
$R$ and $I$ filters respectively. The Cepheids V24 and V26 are located by
about 1.0-1.5 mag below the ridge line in the P-L diagram which makes them
a possible Population II Cepheid variables. Therefore, we did not consider
these two Cepheids for the zero points evaluation. Using remaining 24 Cepheids, 
we found a zero points of 23.54$\pm$0.09  and 23.02$\pm$0.07 mag in $R$ and $I$ 
filters respectively. The dashed envelope lines are drawn 0.5 magnitude from
the fitted line, representing the expected intrinsic scatter of P-L relation
due to the finite width of the instability strip in the H-R diagram (Sandage
1958, Sandage \& Tammann 1968), blending of Cepheids as well as the range of
interstellar extinction.

The above zero points give us a mean apparent distance modulus of
25.12$\pm$0.09 and 24.96$\pm$0.07 mag in $R$ and $I$ filters respectively. This 
yields a mean colour excess of
\begin{equation}
E(R-I)  = (m-M)_{R} - (m-M)_{I}  =  0.16~$mag$
\end{equation}
Using this and the standard extinction law given by Cardelli et al. (1989), we
found a total extinction of 0.63 and 0.47 mag in $R$ and $I$ filters
respectively. Correcting the apparent distance modulus for the total extinction,
we obtain a distance modulus of 24.49$\pm$0.11 mag for the M31 which corresponds
to a distance of 790$\pm$45 Kpc. Stanek \& Garnovich (1998) based on red
clump method and Holland (1998) based on globular cluster CMDs estimated a
distance of M31 as 780 Kpc while a recent determination of Freedman et
al. (2001) based on Cepheid P-L relation suggest a distance of 750 Kpc for M31.
Present estimate of M31 distance is thus agrees well with them.
\begin{figure}[t]
\centering
\includegraphics[height=12cm,width=9.0cm]{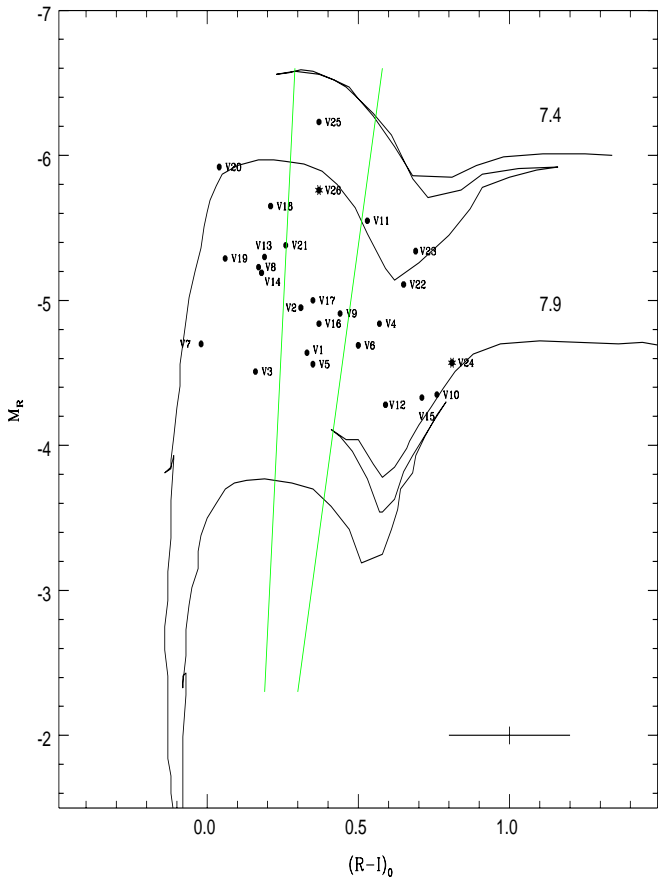}
\caption{The colour-magnitude diagram for the Cepheids under discussion.
Symbols are the same as in Fig. 7. The solar metallicity isochrones by Bertelli
et al. (1994) for logA = 7.4 and 7.9 are also plotted. Average magnitude of
errors in observations are shown in the right bottom corner of the figure.}
\end{figure}
\subsection {The Colour-Magnitude Diagram}
In the M$_R$, $(R-I)_0$ diagram, locations of the Cepheids under discussion
are shown (Fig. 6). The isochrones by Bertelli et al. (1994) for solar
metallicity drawn for logA as 7.4 and 7.9 are also plotted which supports the
age estimated from equation (5). As most of the detected Cepheids ranges
in magnitude from 19 to 21 in $R$ filter and 18 to 20 in $I$ filter,
there can be an error of $\sim$0.2 mag in the colour determination.
\subsection {Location of the Cepheids}
The locations of the Cepheids (see Fig. 3) indicate that most of them are
towards the inner side of M31 (bulge direction), where, possibly, the star
formation was active a few million years ago. This inference is further
supported by the fact that, the early type supergiants detected in our field
($R$ $\sim$ 16 to 19 mag) also occupy the same strip where the Cepheids are
found. So it is tempting to suggest that we are possibly tracing one of the
spiral arms of M31.
\subsection{Brief description of the Cepheids}
We detect 26 Cepheids in which V22 and V24 are newly discovered while V2, V3,
V6, V7, V9, V10, V14, V15, V17, V20 and V23, as reported variables in earlier
studies, are identified Cepheids for the first time. Out of the 26 Cepheids,
only 12 are monitored for a period of four years while remaining are observed
only for last three years. The period of 4 Cepheids were cross-examined using
INT data (obtained with 2.5m telescope at La Palma, Canarie Islands) using
Lafler-Kinman (1965) periodogram which agree quite well with our estimated
period except V8 which have a low amplitude, highly scattered light curve and
needs further investigation. One of the Cepheid, V4, could not be ascribed as
a Cepheid variable purely on the basis of our data. However, $V$ band light
curve of the same star in KAL99 clearly indicates its Cepheid nature. Other
interesting Cepheids are briefly described below:

V19--- KAL99 obtained a period of 16.999 days for this Cepheid while we
obtained a period of 17.83$\pm$0.03 days using 18 data points. The light curve
for this Cepheid is extremely under-sampled and poorly distributed in phase in
our data which could be the reason for discrepancy between two periods. 

V22--- It is discovered in the present study with a period of 26.96$\pm$0.04
days. A small bump in the falling branch can be seen in both filters.

V24--- This is another Cepheid discovered by us. It has a period of
35.12$\pm$0.05 days and an exceptionally large value of $(R-I)$ colour of 0.98
mag. In the P-L diagram, this Cepheid deviates maximum towards the fainter
side in both filters and could be a possible Population II Cepheid.

V26--- This is the longest period Cepheid in our sample. We obtained its
period as 56.02$\pm$0.08 days in agreement with a period of 56.116 days
derived by KAL99. In the P-L diagram, this Cepheid is also located downwards
in both filters and it could also be a Population II type Cepheid.
\section {Other variables}
In addition to the Cepheid variables reported in our field, 333 other variable
stars are also found in the present study. Most of them appear to be irregular
variables. Only 115 stars are found periodic for which we could determine
approximate period and magnitude. In Table 5, available in the electronic form
either from authors or at the CDS, we have given the identification number,
celestial coordinates, mean magnitude in $R$ and $I$ filters, amplitude
variation and nature of these 333 stars in $R$ filter. Out of these 333 stars,
32 were already categorized (see Table 5). V176 and V284 are suspected as a
binary and light curve for one of them is shown in Fig 7(a). We have given the
light curve of a periodic star V154 with period $\sim$236 days in Fig. 7(b)
while the light curve of an irregular variable V252 is shown in Fig 7(c). 
\begin{figure}[h]
\centering
\includegraphics[height=11.0cm,width=9.0cm]{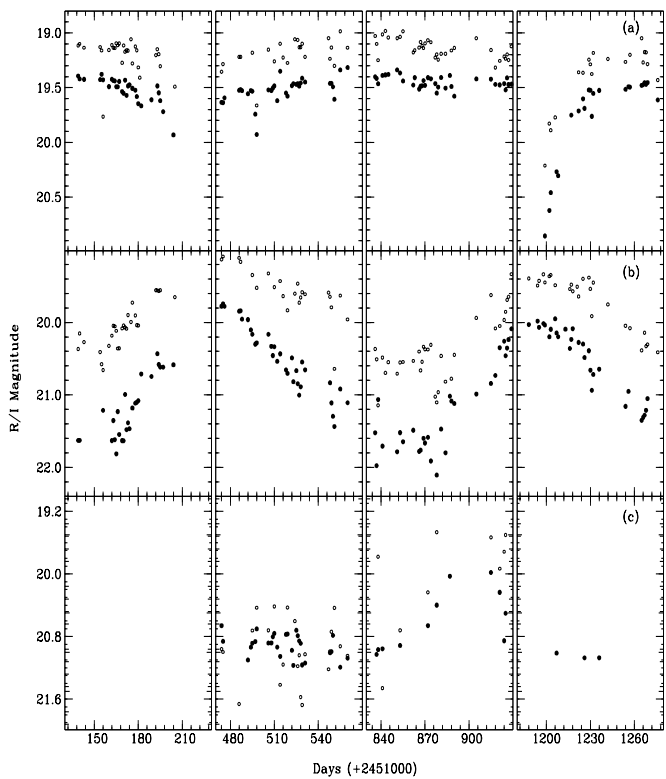}
\caption{The typical light curve of 3 other variables (for further detail see
Sect. 6). Filled and open circles represent $R$ and $I$ filters respectively.}
\end{figure}
\section{Conclusion}
We have been imaged $\sim13 \times 13$ arcmin$^2$ region of the M31 disk for
about 4 years during 1998 to 2001 in $R$ and $I$ passbands. Using 141 nights
of observation spanning a period of $\sim$1200 days, we have detected 26
Cepheids among which variability in 2 stars has been found for the first time.
We have established the Cepheid nature of 11 variables in the sample, while we
confirm the period and nature of variability in the other 13 stars. The long
duration data enabled us to obtain accurate period and mean magnitude of the
Cepheids. The period obtained for these Cepheids ranges from $\sim$7.5 to 56
days. They range in age from 22 Myrs to 75 Myrs and occupy a strip where the
early type supergiants too are located. Using the period-luminosity relation of 
the Cepheids, we derive a distance of 790$\pm$45 Kpc for M31. There could be
additional uncertainty in the distance determination due to blending effect in
Cepheids as well as variable extinction within the observed region.

In addition to Cepheids, we also detect 333 variable stars. Out of them, 115
are periodic variables, 2 suspected eclipsing binaries and remaining 216 stars
appear to be irregular variables.

We have thus demonstrated once more that a meter class telescopes like ours can
play important role in the study of variable stars provided large amount of
telescope observing time is made available. These data are also valuable for
the search of gravitational microlensing events towards M31. We are in the
process of identifying lensing candidates in our target field and their
detailed study shall be presented in the forthcoming publication.

~

{\it Acknowledgments}
We would like to thank the referee Dr. B.J.Mochejska for useful remarks on the
manuscript. We are also thankful to Dr. Vijay Mohan for his helpful advice
during observations and Drs. Jean Kaplan, Alan Bouque and Vincenzo Cardone
for useful discussions during the course of the work. This study is a
part of the project 2404-3 supported by Indo-French center for the Promotion
of Advance Research, New Delhi.

\end{document}